\documentstyle[aps,twocolumn]{revtex}
\begin{document}
\preprint{USACH-FM-00-06}
\title{ Noncommutativity and the Aharonov-Bohm Effect}
\author{ J. Gamboa$^1$\thanks{E-mail: jgamboa@lauca.usach.cl}, M. Loewe$^{2}$
\thanks{E-mail:
mloewe@fis.puc.cl,} and
J. C. Rojas$^2$\thanks{E-mail: jrojas@lauca.usach.cl}}
\address{$^1$Departamento de F\'{\i}sica, Universidad de Santiago de Chile,
Casilla 307, Santiago 2, Chile\\
$^2$ Facultad  de F\'{\i}sica, Pontificia Universidad  Cat\'olica de Chile,
Casilla 306, Santiago 22, Chile}
\maketitle
\begin{abstract}
The possibility of detecting noncommutive space relics is analyzed by using the Aharonov-Bohm effect. If
space is non-commutative, it turns out that  the holonomy receives kinematical corrections that tend to
diffuse the fringe pattern. This fringe pattern has a non-trivial energy dependence and, therefore, one could
observe noncommutative effects by modifying the energy of the incident electrons beam in the  Tonomura
experimental arrangement.
\end{abstract}
\section{Introduction}
There are arguments in string theory suggesting that  spacetime could be non-commutative\cite{string}.
Although this property
might be an argument in favor of new renormalizables effective field theories\cite{various}, it represents also
a trouble because we
need to explain the transition between the commutative and noncommutative regimes.

If the noncommutative effects are important at very high energies, then one could posit a decoupling
theorem that produces the standard quantum field theory as an effective field theory and that does not remind
the noncommutative effects. However, our experience in atomic and molecular physics\cite{mol} strongly
suggests that the decoupling is never complete, and the high energy effects appear in the  effective action as
topological remnants\cite{fw}.

Following this idea we would like to consider an example, related to topological aspects, where the
appearance of noncommutative effects could be relevant. A natural candidate is the Aharonov-Bohm effect
\cite{ab}, where we know that the line spectrum does not depend on the relativistic nature of the electrons
\cite{vitorio}. However, as will see below, the spatial noncommutativity smears the spectral lines and this
effect, in principle, could be observed by increasing the energy of the electrons in the Aharonov-Bohm
experiment.

The article is organized as follows; in section 2, we discuss the noncommutative Aharonov-Bohm effect
following a very simple perturbative approach and a generalized formula for the holonomy is given. In section
3, we discuss the physical consequences of our result and we argue how to observe noncommutative effects
from the Tonomura experiment. Finally, an appendix discussing the fringe pattern in the relativistic Aharonov-Bohm effect is also considered.

\section{The Noncommutative Aharonov-Bohm Effect}

Let us start considering a spinless electron moving on a two dimensional noncommutative plane
$\Re^2 - \{0\}$, where $\{0\}$ is constructed putting a solenoid with a magnetic field concentrated along the
$x_3$ axis. Since in the noncommutative case space is a collection of cells, the reader will notice that this
rearrangement is just an approximation valid for small values of the anticommutative $\theta$ parameter.

The field tensor in the noncommutative plane is
\begin{equation}
{\hat F}_{\mu \nu} = \partial_\mu A_\nu - \partial_\nu A_\mu +
i e A_\mu \star A_\nu - i e A_\nu \star A_\mu, \label{1}
\end{equation}
where $\star$ is the Moyal product defined as
 \begin{equation}
 {\bf A}\star {\bf B}({\bf x}) = e^{\frac{i}{2}\theta^{ij}\partial^{(1)}_i \partial^{(2)}_j}
 {\bf A}({\bf x}_1) {\bf B}({\bf x}_2) \vert_{{\bf x}_1={\bf x}_2={\bf x}}, \label{moyal}
\end{equation}
and we will use $\theta^{ij}= \epsilon^{ij} \theta$.

As we are assuming that the noncommutative effects are small, one can expand the Moyal product retaining
only the linear term in $\theta$,  {\it i.e.}
\begin{equation}
{\hat F}_{\mu \nu} = \partial_\mu A_\nu - \partial_\nu A_\mu +
e \theta \epsilon^{\alpha \beta} \partial_\alpha A_\mu \partial_\beta
A_\nu. \label{2}
\end{equation}

Then, we must construct a gauge potential such that the magnetic field  $B$ vanishes everywhere except at the
origin. In order to do that, we proceed as in the commutative case, {\it i.e.} from
\begin{equation}
{\bf A} = \frac{ -x_2 {\hat x}_1 + x_1 {\hat x}_2}{x_1^2 + x_2^2}.\frac{\phi}{2 \pi} \label{3}
\end{equation}
 where $\phi$ is the flux inside  the solenoid. We construct the non-commutative potential by means of the
 Ansatz
\begin{eqnarray}
A_1 &=& - x_2 f(r^2).\frac{\phi}{2 \pi}, \nonumber
\\
A_2 &=& x_1 f(r^2).\frac{\phi}{2 \pi}. \label{4}
\end{eqnarray}

Since we have taken an expansion of the Moyal product in eq. (\ref{2}), the ordinary product in (\ref{4}) must
be understood.  Then, as in the commutative case, we impose $B_3 = {\hat F}_{12}=0 $ outside  the
solenoid implying the
condition
\begin{equation}
 2 f + 2 r^2 f^{'} - e \theta ( f^2 + 2 r^2 f f^{'}) = 0, \label{5}
\end{equation}
where $f^{'} = df/dr^2$.

This differential equation can be integrated easily and yields the following solution
\begin{eqnarray}
f &=&  \frac{1}{ e \theta} \pm \frac{1}{ e \theta} \sqrt{1 - \frac{e\theta}{r^2}} \nonumber
\\
&=& \frac{1}{e \theta} \pm  \frac{1}{ e \theta} [ 1 -
\frac{e\theta}{2r^2} - \frac{ e^2 \theta^2}{8r^4} + ...]. \label{6}
\end{eqnarray}

From (\ref{6}) we see that the commutative limit is smooth for the  minus sign
in the above equation {\it i.e.}
\begin{equation}
 f = \frac{1}{2r^2} + \frac{e \theta}{8r^4} + ...\label{7}
\end{equation}
and the potential
\begin{eqnarray}
A_1 &=& - x_2 \biggl( \frac{1}{2(x_1^2 + x_2^2)} +
\frac{e \theta}{8(x_1^2 + x_2^2)^2} + ...\biggr).\frac{\phi}{2 \pi}\nonumber
\\
A_2 &=&  x_1 \biggl( \frac{1}{2(x_1^2 + x_2^2)} +
\frac{e \theta}{8(x_1^2 + x_2^2)^2}+ ...\biggr).\frac{\phi}{2 \pi}, \label{8}
\end{eqnarray}
describes a magnetic field zero everywhere except at the origin. This potential is also the non-
commutative generalization of the
magnetic monopole for $\theta$ small\cite{gross}.

Next step is to solve the Schr\"odinger equation for a particle with mass $m$ moving in the field (\ref{8}).
However, instead of
doing this we remind that, in the commutative case, the  Schr\"odinger equation in an external gauge
potential is solved by
\begin{equation}
\psi = e^{ie\int_C dx^j A_j} \varphi, \label{9}
\end{equation}
where the $U(1)$ holonomy $ e^{ie\int_C dx^j A_j}$, in general,  is a non-integrable factor,
{\it i.e.} dependent
on the path $C$,
being $\varphi$
the free solution of the Schr\"odinger equation.

However,  although formally (\ref{9}) solve the Schr\"odinger equation, the holonomy  involves in a non trivial
way the dynamics
of the gauge potential and it hides all the complications of ${\bf A}$. Our goal below will be to find an
approximate expression
for the holonomy for $\theta$ small.

Let us suppose that the operator
$D_j = -i \partial_j + e A_j$ satisfies the eigenvalue equation
\begin{equation}
D_j \star \psi = k_j \psi. \label{10}
\end{equation}

Then, using (\ref{10})
the Schr\"odinger equation becomes
\begin{equation}
{\hat H} \psi = \frac{1}{2m} D_j \ast D_j \ast \psi =
\frac{1}{2m}k_j k_j \psi. \label{11}
\end{equation}

In order to solve (\ref{10}) we use the Ansatz
\begin{equation}
\psi = e^{F}, \label{12}
\end{equation}
and therefore, for $\theta$ small
\begin{eqnarray}
D_j \psi&=& -i\partial_j e^F + e A_j \ast e^F \nonumber
\\
&= & e^F [ -i \partial_j F + e A_j + \frac{i}{2} e \theta \epsilon^{lm}
(\partial_l A_j)(\partial_m F) ] \nonumber
\end{eqnarray}
and furthermore
\begin{equation}
-i \partial_j F + e A_j + \frac{i}{2} e \theta \epsilon^{lm} (\partial_l A_j)
(\partial_m F) = k_j. \label{13}
    \end{equation}

Now, one can solve (\ref{13}) perturbatively by expanding  $F$ and $A_j$ in powers of $\theta$,
{\it i.e.}
\begin{eqnarray}
F &=& F^{(0)} + \theta F^{(1)} + ... \label{14}
\\
A_j &=& A^{(0)}_j + \theta A_j^{(1)} + ..., \label{144}
\end{eqnarray}
then at zero order in $\theta$, equation (\ref{13}) gives
\begin{equation}
-i \partial_j F^{(0)} + e A^{(0)}_j = k_j, \label{15}
\end{equation}
and the following expression for $F^{(0)}$ is obtained
\begin{equation}
 F^{(0)} = ik_j (x-x_0)_j - ie \int_{x_0}^x dx_j A^{(0)}_j. \label{16}
 \end{equation}

The first term in the RHS is just the free particle solution if we interpret $k_j$ as the wave number and the
second term is the $U(1)$ holonomy for the commutative case. Thus, at zero order, we reproduce the
commutative solution of the Schr\"odinger equation.

If we retain first order terms in $\theta$, the following  differential equation is obtained
\begin{equation}
-i\partial_j F^{(1)} + eA^{(1)}_j + \frac{i}{2} e \theta \epsilon^{lm} (\partial_l  A^{(0)}_j)
(\partial_m F^{(0)}) = 0, \label{17}
\end{equation}
and the integration of (\ref{17}) gives
\begin{equation}
 F^{(1)} = -e \int_{x_0}^x dx_j A^{(1)}_j - \frac{ie \theta}{2}  \int_{x_0}^x dx_j \epsilon^{ml} ( k_m - e A^{(0)}_m)
\partial_l A^{(0)}_j ). \label{18}
 \end{equation}

The first term in the RHS is an additive correction to the commutative holonomy, {\it i.e.}
\begin{equation}
-e \int_{x_0}^x dx_j ( A^{(0)}_j + \theta A^{(1)}_j) =: -e \int_{x_0}^x dx_j  A_j, \nonumber
\end{equation}
and a second one include a velocity dependent term  which  can be written as follows\footnote{A different contribution was
obtained in \cite{chaichian}. However, the mistake of these authors was corrected after our paper appeared in a the subsequent version of \cite{chaichian}. We would like to thank to the referee for drawing our attention on this point.}:
\begin{equation}
- \frac{i}{2} e \int_{x_0}^x dx_j \,\,\epsilon^{ml} k_m \partial_l A^{(0)}_j = -\frac{i}{2}em
\int \,dx_j \, ({\bf v} \times
\nabla A^{(0)}_j)_3, \label{19}
\end{equation}

For the third term our calculation yields to
\begin{equation}
\int_{x_0}^x dx_j \,({\bf A}^{(0)}\times \nabla A^{(0)}_j)_3.\nonumber
\end{equation}

Thus,  at this order in $\theta$, the nonconmutative holonomy is
\begin{eqnarray}
{\cal W}(x,&x_0&) =\nonumber
\\
&=&\exp\biggl[- ie \int_{x_0}^x dx_j A_j  -
    \frac{i}{2} m e \theta \int_{x_0}^x dx_j \,[({\bf v}\times \nabla A^{(0)}_j)_3  \nonumber
    \\
    &-& e
({\bf A^{(0)}}\times \nabla A^{(0)}_j)_3 ]\biggr]. \label{21}
\end{eqnarray}

Now, we analyze the different terms in (\ref{21}); the first one in the exponential is the usual holonomy
(corrected at order $\theta$) which classifies the differents homotopy classes.

The term
\begin{equation}
\int_{x_0}^x dx_j \,\,[{\bf A^{(0)}}\times \nabla A^{(0)}_j]_3, \label{22}
\end{equation}
is a non-commutative correction to the vortex decaying as $1/r^3$ and does not contribute to the line
spectrum.

Finally, the term
\begin{equation}
\theta \int_{x_0}^x dx_j \,[({\bf v}\times \nabla A^{(0)}_j]_3 \label{23}
\end{equation}
is a velocity dependent correction insensitive to the topology of the manifold.  If ${\bf v}$
increases,  the contribution of
(\ref{23}) to the holonomy oscillates very quickly smearing the fringe pattern.
The above result is completly different to the commutative Aharonov-Bohm effect where the fringe pattern is
insensitive to the relativistic nature of the electrons \cite{vitorio}.

\section{Possible experimental strategy for detecting noncommutative effects}

In this section we will analyse the possibility of detecting spatial noncommutative effects by using the Tonomura experiment. Our claim is that if one increase the energy of the incident electrons  beam, then  one should observe a squeezing  of the fringes pattern. As we expect that  $\theta$ is small, then  $\theta$ --as was assumed in the previous section--  is a perturbative parameter and the main part of the fringe pattern is contained in the commutative sector of the Aharonov-Bohm. 

If we now modify the energy of the electrons beam, then  we should see noncommutative corrections.  In the original Tonomura experiment
\cite{tono}, they consider incident beams with energies among 80, 100 and 125 keV but they did not find changes in the fringes pattern. In our opinion, this is explained by the perturbative stability of the Aharonov-Bohm fringes pattern but, if we can consider incident energies among 160-500 keV (or higher) spectral differences should be observed.  At this energies the effects of the penetrability can be important, but this fact does not affect the flux-dependent phase difference\cite{ola}.

Presently, we have several bounds for  $\theta$\cite{cotas}. The noncommutative contributions will, in general, be tiny. We hope, however, that futureTonomura or improved Tonomura kind experiments will be able to distinguish these new effects.

\acknowledgments
We would like to thank to G. V. Dunne F. M\'endez and V. O. Rivelles for several discussions. We would like to thank also Professor Akira Tonomura by inform us about energy of the incident electron beam. This work has
been partially supported by the grants Nr. 1010596 and Nr. 1010976 from Fondecyt-Chile and Dicyt-USACH.

\section{Appendix}

In this appendix, we would like to discuss some implications of the relativistic Aharonov-Bohm effect. In particular, we would like to emphasize that due the topological character of the Aharanov-Bohm effect (and also to the fact that the radius of the solenoid is zero)  the 
interference pattern does not change by relativistic corrections. 

This last fact can be seen as follows; using the reference \cite{vitorio}
one see that the Green function associated to the usual Aharonov-Bohm effect is given by
\begin{equation}
G[x,x^{'}] = \sum_{n=-\infty}^{\infty} (-i)^{|n+\phi|} \exp[-i(n+\phi)] F_{|n+\phi|}, \label{gre1}
\end{equation}
where $\phi$ is the magnetic flux and the function $F_{|n+\phi|}$  for the non-relativistic case is
\begin{equation}
F_{|n+\phi |}= \frac{m}{2\pi i} \exp[ \frac{2m i}{\tau} (r^2 + r^{'2})] J_{|n+\phi|} ( \frac{mrr^{'}}{\tau}), \label{nr}
\end{equation}
where $\tau = t-t^{'}$ and $J_{\alpha }$ are Bessel functions.
For the relativistic case the calculation is similar.
Indeed,  after using the proper-time gauge the function
$F_{|n+\phi |}$ becomes
\begin{eqnarray}
F_{|n+\phi |}&=& \nonumber
\\
\int d^2p \int_0^\infty d\lambda \exp[&i& p_\mu \Delta x^\mu - \frac{\lambda}{2} (p^2 +m^2)] J_{|n+\phi|} (
\frac{r r^{'}}{\lambda}). \label{r}
\end{eqnarray}
where $\lambda= N(0) (t-t^{'})$ with $N(0)$ the einbein.

If we use the Poisson summation formula, then in both the relativistic as well as
in the non relativistic case, the Green function is
\begin{equation}
G[x,x^{'}] = \sum_{n=-\infty}^{\infty}  e^{2i\pi n \phi} K_n, \label{ho}
\end{equation}
where $K_n$ is defined as
\begin{equation}
K_n = \int_{-\infty}^\infty d\omega\,(-i)^{|\omega|}\, e^{-i\omega \phi}\, F_{|\omega|}, \nonumber
\end{equation}
and, as a consequence, the wave function becomes
\begin{equation}
\psi (x) = \sum_{n=-\infty}^{\infty} e^{2i\pi n \phi} \varphi_n (x), \label{wf}
\end{equation}
with
\begin{equation}
\varphi_n (x) = \int \,dy\, \,G_n\,[\,x,y]\, \psi (y),
\end{equation}
being $\varphi_n$ and $G_n [\,x,y]$, respectively, the wave and Green functions for
the n-th homotopy class \cite{dewitt}.

Thus, from (\ref{wf}) one see that the relativistic character of the system is contained in $K_n$ and
only  the exponential factor, which does not depend on the energy,
is responsible for the fringe pattern. This result reflects the topological nature of
the commutative Aharonov-Bohm effect. However, our formula (\ref{23}) show us that the noncommutative
Aharonov-Bohm effect is  radically different  because the fringe pattern must change when the
electrons are getting higher energies.

\end{document}